\documentclass[journal]{IEEEtran}
\usepackage[dvips]{graphicx}

\begin{document}
\title{An Analytical Formulation of Power System Oscillation Frequency}
\author{
        Bin~Wang,~\IEEEmembership{Student Member,~IEEE,}
        and~Kai~Sun*,~\IEEEmembership{Senior Member,~IEEE}
          \thanks{* K.~Sun is with Department of EECS, University of Tennessee, Knoxville, TN 37996 (e-mails:  kaisun@utk.edu).}
}
\maketitle

\begin{abstract}
This letter proposes an analytical approach to formulate the power system oscillation frequency under a large disturbance. A fact is revealed that the oscillation frequency is only the function of the oscillation amplitude when the system's model and operating condition are fixed. Case studies also show that this function is damping-insensitive and could be applied to an inter-area model of a multi-machine power system.
\end{abstract}

\section{Introduction}
\IEEEPARstart{P}{ower} systems' electromechanical oscillations are detrimental to system stability. A traditional small signal analysis approach linearizes a system's nonlinear differential-algebraic equation model near a specific operating condition to estimate its natural oscillation frequency (OF) but neglect the nonlinearity of the system, so the natural oscillation frequency cannot describe the accurate OF, especially under a large disturbance. When estimating the OF from power system measurements, most signal processing based methods, e.g. Fourier Analysis and Prony Analysis, need data over a time window much longer than the period of the targeted oscillation mode to estimate an average OF while ignoring any instantaneous change in the window. For a pendulum system, it is feasible to analytically formulate its period or frequency under a large disturbance by elliptic integral of the first kind\cite{website} but there is no similar study for power systems, even for a single-machine-infinite-bus (SMIB) or a two-area power system, which can be regarded as a fictitious damped pendulum system with a constant torque.

This letter will analytically formulate the power system's OF under a large disturbance and introduce the Frequency-Amplitude curve as a characteristic in nature with the power system about a specific oscillation mode. The formulation will then be tested on a SMIB system and a two-area system.

\section{Oscillation Frequency Formulation}
Considering the motion equation
\begin{equation} \label{eq:swing}
\left\{ \begin{array}{ll}
\Delta\dot{\delta}=\omega_{0}\Delta\omega_{r} \\
\Delta\dot{\omega}_{r}=(P_{m}-P_{e}-D\Delta\omega_{r})/2H
\end{array} \right.
\end{equation}
$\omega_{0}$ is the synchronous frequency, $\Delta\delta$ is the rotor angle deviation relative to its steady-state value $\delta_{0}$, $\Delta\omega_{r}$ is the per-unit rotor speed deviation, $P_{m}=P_{max}\sin\delta_{0}$ and $P_{e}=P_{max}\sin(\delta_{0}+\Delta\delta)$ represent the per-unit mechanic and electric torques, respectively, where $P_{max}$ is the steady-state maximum power transfer, and $H$ and $D$ represent the inertia and damping factor of the machine, respectively. For simplicity, assume $P_{m}$ is constant under the disturbance. Eliminating $\Delta\omega_{r}$ leads to (\ref{eq:de1}), where $\beta=P_{max}\omega_{0}/2H$ for simplicity.
\begin{equation} \label{eq:de1}
\Delta\ddot{\delta}+(D/2H)\Delta\dot{\delta}+\beta\sin(\delta_{0}+\Delta\delta)=\beta\sin\delta_{0}
\end{equation}

Unlike a pendulum system whose oscillation is symmetric about the equilibrium, a SMIB system has no such symmetry\cite{pendulum}: for each cycle of its oscillation, the motions in the upper and lower halves of cycle are different and hence should be formulated separately. Assume that the maximum and minimum angle deviations, $\Delta\delta_{max}$ and $\Delta\delta_{min}$, are known and the damping is zero.

The upper half of cycle is the time taken from the equilibrium to the maximum plus the returning time. Only the upswing is considered according to the symmetry of oscillation within this half cycle. Define potential energy at the equilibrium to be zero. The kinetic and potential energies are calculated by (\ref{eq:energy}). From the law of conservation of energy, the total energy holds from $\Delta\delta$ to $\Delta\delta_{max}$ during the upswing. Solve for $\textrm{d}t$ and get (\ref{eq:dt}).
\begin{equation} \label{eq:energy}
\left\{ \begin{array}{ll}
E_{k}=(\textrm{d}\Delta\delta/\textrm{d}t)^{2}H/\omega_{0} \\
E_{p}=P_{max}\bigr(\cos\delta_{0}-\cos(\delta_{0}+\Delta\delta)-\Delta\delta\sin\delta_{0}\bigr)
\end{array} \right.
\end{equation}
\begin{eqnarray} \label{eq:dt}
\textrm{d}t=\bigl[2\beta\bigl(\cos(\Delta\delta+\delta_{0})-\cos(\Delta\delta_{max}+\delta_{0})\nonumber\\
+(\Delta\delta-\Delta\delta_{max})\sin\delta_{0}\bigl)\bigr]^{-\frac{1}{2}}\textrm{d}\Delta\delta
\end{eqnarray}

By integrating both sides for $\textrm{d}\Delta\delta$ from 0 to $\Delta\delta_{max}$, the left side is the total time of that motion, doubling which gives the time of the upper half of cycle (\ref{eq:tu}). Next, change the integral variable by (\ref{eq:trans1}), where $k=\sin(\Delta\delta_{max}/2)$. Then we have (\ref{eq:trans2}).
\begin{eqnarray} \label{eq:tu}
T_{u}=2\int_{0}^{\Delta\delta_{max}}\bigl[2\beta\bigl(\cos(\Delta\delta+\delta_{0})-\cos(\Delta\delta_{max}\nonumber\\
+\delta_{0})+(\Delta\delta-\Delta\delta_{max})\sin\delta_{0}\bigl)\bigr]^{-\frac{1}{2}}\textrm{d}\Delta\delta
\end{eqnarray}
\begin{equation} \label{eq:trans1}
\sin\varphi=\sin(\Delta\delta/2)/\sin(\Delta\delta_{max}/2)
\end{equation}
\begin{equation} \label{eq:trans2}
\left\{ \begin{array}{ll}
\Delta\delta=2\arcsin(k\sin\varphi) \\
\textrm{d}\Delta\delta=2k\cos\varphi({1-k^{2}\sin^{2}\varphi)^{-\frac{1}{2}}}\textrm{d}\varphi
\end{array} \right.
\end{equation}

Note that when $\Delta\delta$ increases from 0 to $\Delta\delta_{max}$, $\varphi$ increases from 0 to $\pi/2$. Thus, (\ref{eq:tu}) becomes
\begin{eqnarray} \label{eq:tu2}
T_{u}=2\int_{0}^{\pi/2}g_{u}(\sin\varphi)\frac{2k\cos\varphi}{\sqrt{1-k^{2}\sin^{2}\varphi}}\textrm{d}\varphi
\end{eqnarray}
\begin{eqnarray} \label{eq:gu}
g_{u}=\bigl[2\beta\bigl(\cos(2\arcsin(k\sin\varphi)+\delta_{0})-\cos(\Delta\delta_{max}\nonumber\\
+\delta_{0})+(2\arcsin(k\sin\varphi)-\Delta\delta_{max})\sin\delta_{0}\bigl)\bigr]^{-\frac{1}{2}}
\end{eqnarray}

Similarly, $T_{l}$ will be
\begin{eqnarray} \label{eq:tl}
T_{l}=2\int_{0}^{-\pi/2}-g_{l}(\sin\varphi)\frac{2k\cos\varphi}{\sqrt{1-k^{2}\sin^{2}\varphi}}\textrm{d}\varphi
\end{eqnarray}
where $g_{l}$ has the same form as $g_{u}$ except that $\Delta\delta_{max}$ is replaced by $\Delta\delta_{min}$ and $k=-\sin(\Delta\delta_{min}/2)$. Then, define OF $f$ as (\ref{eq:of}).
\begin{equation} \label{eq:of}
f=1/(T_{u}+T_{l})
\end{equation}

When $\Delta\delta_{max}$ and $\Delta\delta_{min}$ approach 0, the general OF defined by (\ref{eq:of}) will degrade to the nature frequency of the linearized model of the power system. Here is
\begin{equation} \label{eq:lmtf}
\lim_{\Delta\delta_{max},\Delta\delta_{min} \to 0}f=\sqrt{\beta\cos \delta_{0}}/2\pi
\end{equation}

The following two methods can either solve $\Delta\delta_{max}$ and $\Delta\delta_{min}$ directly from the model or obtain their values from measurement data:

Assume initial values of (\ref{eq:de1}) to be $\Delta\delta(0)$ and $\Delta\dot{\delta}(0)$. Because $E_{k}=0$ at each extreme point $\Delta\delta_{ep}$ (either $\Delta\delta_{max}$ or $\Delta\delta_{min}$), the conservation of energy gives (\ref{eq:ep}) about $\Delta\delta_{ep}$. When the system is stable, (\ref{eq:ep}) has one positive root and one negative root, i.e. $\Delta\delta_{max}$ and $\Delta\delta_{min}$, respectively.
\begin{eqnarray} \label{eq:ep}
(\Delta\dot{\delta}(0))^{2}+\beta\bigr(\cos(\Delta\delta_{ep}+\delta_{0})-\cos(\Delta\delta(0)+\delta_{0})\nonumber\\
+(\Delta\delta_{ep}-\Delta\delta(0))\sin\delta_{0}\bigr)=0
\end{eqnarray}

Alternatively, if the time series of $\delta$ are measured, by subtracting an estimate of its steady-state value, we may obtain values of $\Delta\delta_{max}$ and $\Delta\delta_{min}$.

Finally, two approaches can be employed to calculate the OF in (\ref{eq:of}). The first approach uses power series to approximate (\ref{eq:tu2}) and (\ref{eq:tl}) which gives the OF by the sum of the first $N$ terms of the series as shown in (\ref{eq:psfg}), where $f_{u}$ and $f_{l}$ are integrands of (\ref{eq:tu2}) and (\ref{eq:tl}), respectively. When only keeping the first term, the approximation of OF is shown in (\ref{eq:psf}). The second approach is to estimate (\ref{eq:tu2}) and (\ref{eq:tl}) numerically.
\begin{equation} \label{eq:psfg}
f_{ps}=1/\sum_{i=0}^{N-1}\int_{0}^{\pi/2}\frac{f_{u}^{(i)}(0)\sin^{i}\varphi+f_{l}^{(i)}(0)\sin^{i}(-\varphi)}{i!}\textrm{d}\varphi
\end{equation}
\begin{equation} \label{eq:psf}
f_{ps}|_{N=1}=\frac{\sqrt{\beta}}{\sqrt{2}\pi}\Bigl(\frac{\sin(\Delta\delta_{max}/2)}{m(\Delta\delta_{max})}
+\frac{\sin(-\Delta\delta_{min}/2)}{m(\Delta\delta_{min})}\Bigr)^{-1}
\end{equation}
\begin{equation} \label{eq:m}
m(x)=\sqrt{\cos \delta_{0}-\cos(\delta_{0}+x)-x\sin \delta_{0}}
\end{equation}

Define the oscillation amplitude (OA) as $\frac{(\Delta\delta_{max}-\Delta\delta_{min})}{2}$. Equation (\ref{eq:psfg}) discovers that OF of a specific mode is only the function of OA when the system’s model and operating condition are fixed. Thus, a Frequency-Amplitude (F-A) curve can be drawn from (\ref{eq:psfg}) as a characteristic of the system. The OA when OF approaches zero indicates the stability limit.

\section{Case Study}
First, the influence of damping is illustrated on a SMIB power system, and then the proposed general OF is validated on Kundur's two-area power system\cite{kundur}.

\subsection{Test on a SMIB Power System}
Let $H=3$, $\omega_{0}=120\pi$, $P_{max}=1.3$ and $\delta_{0}=0.8$ in (\ref{eq:de1}). Consider three cases: 1) $D=1$, $\Delta\delta(0)=30^{\circ}$ and $\Delta\dot{\delta}(0)=2$rad/s; 2) $D=1$, $\Delta\delta(0)=60^{\circ}$ and $\Delta\dot{\delta}(0)=2$rad/s; 3) $D=3$, $\Delta\delta(0)=60^{\circ}$ and $\Delta\dot{\delta}(0)=2$rad/s. In each case, solve the trajectory of $\Delta\delta$ using the Runge-Kutta method, and for any adjacent two extreme points on the trajectory, their differences in time and angle respectively give estimates for the OF and OA. The formulation (\ref{eq:of}) is calculated numerically by Simpson's rule.

Fig.\ref{fig:smib} gives the F-A curves about 3 cases and the theoretical F-A curve from the formulation, which all curves match well. Comparison between cases 1 and 2 indicates that the initial value does not change the F-A curve. Comparison between cases 2, 3 and the formulation shows that the F-A curve is insensitive to damping.

\subsection{Test on the Kundur's System}
\begin{figure}[htbp]
\begin{minipage}[t]{0.49\linewidth}
\raggedleft
\includegraphics[width=1.55in]{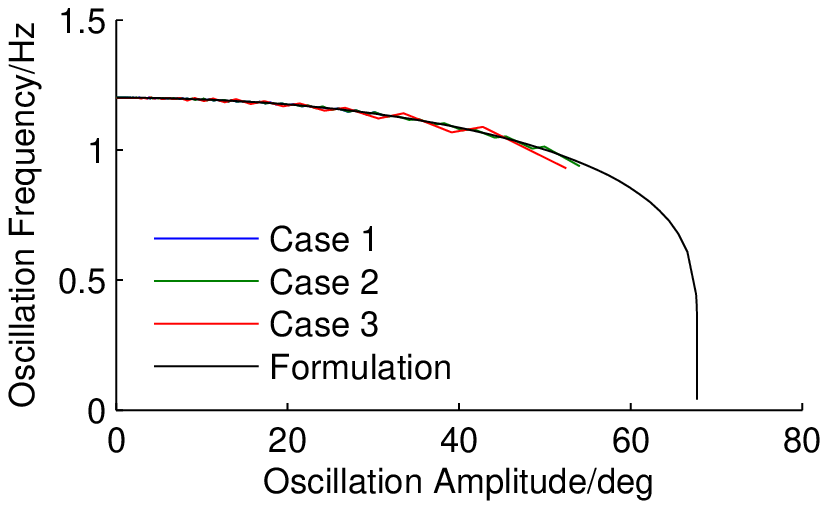}
\caption{Results on a SMIB} \label{fig:smib}
\end{minipage}
\begin{minipage}[t]{0.49\linewidth}
\raggedright
\includegraphics[width=1.55in]{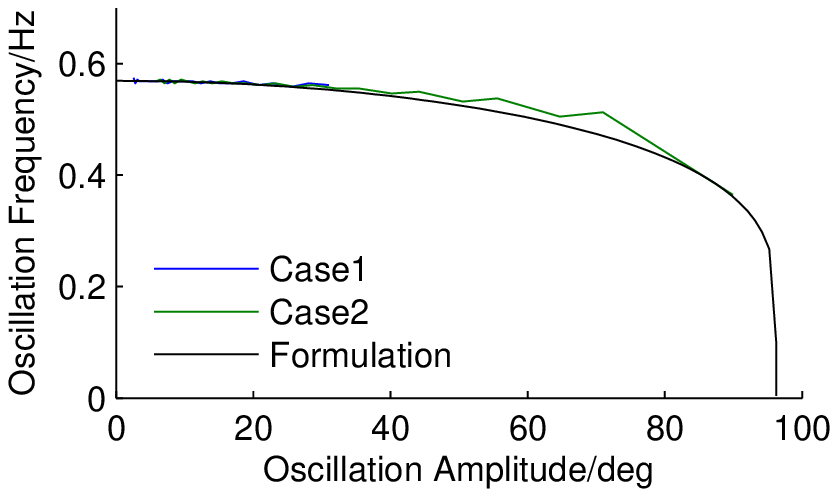}
\caption{Results on Kundur's system} \label{fig:kundur}
\end{minipage}
\end{figure}

Only the inter-area mode is considered and a SMIB equivalent of the system is considered by defining the equivalent angle difference as (\ref{eq:ad}). Case 1 adds a three-phase fault on bus 7 and clears it after 10 cycles by tripping three lines between buses 7 and 8. Case 2 extends the fault duration to 19.9 cycles to approach the transient stability limit. Fig.\ref{fig:kundur} indicates that the proposed formulation could also be applied to a multi-machine power system.
\begin{equation} \label{eq:ad}
\Delta\delta=\frac{\delta_{1}H_{1}+\delta_{2}H_{2}}{H_{1}+H_{2}}-\frac{\delta_{3}H_{3}+\delta_{4}H_{4}}{H_{3}+H_{4}}
\end{equation}

\section{Conclusion}
Analytically formulation for oscillation frequency of a power system has been proposed as a function of the oscillation amplitude when the system's model and operating condition are fixed. The Frequency-Amplitude curve is introduced to characterize the system's nonlinearity with an oscillation mode. The case study indicates the oscillation frequency is insensitive to the damping of the system, and the formulation may also be applied to an inter-area model of a multi-machine power system.

\end{document}